\documentclass[aps,jcp,twocolumn,superscriptaddress,floatfix,10pt]{revtex4}

\usepackage{graphicx}
\usepackage{amsmath}
\usepackage{color}
\usepackage{xcolor}
\usepackage{dcolumn}
\usepackage{soul}
\usepackage{comment}
\usepackage{appendix}
\usepackage{hyperref}
\usepackage[capitalize]{cleveref}
\usepackage[thinc]{esdiff}

\begin{document}
\title{Adhesion study at the interface of PDMS-elastomer and borosilicate glass-slide}

\author{Susheel Kumar}
\affiliation{Department of Mechanical Engineering, Indian Institute of Technology Kanpur, Kanpur UP 208016 India}
\author{Chiranjit Majhi}
\affiliation{Department of Physics, Indian Institute of Technology Kanpur, Kanpur UP 208016 India}
\author{Krishnacharya Khare}
\affiliation{Department of Physics, Indian Institute of Technology Kanpur, Kanpur UP 208016 India}
\author{Manjesh Kumar Singh}
\email[]{manjesh@iitk.ac.in}
\affiliation{Department of Mechanical Engineering, Indian Institute of Technology Kanpur, Kanpur UP 208016 India}

\date{\today}

\begin{abstract}
Adhesion control at the interface of two surfaces is crucial in many applications. Examples are the
design of micro and nanodevices such as microfluidic devices, biochips, and electronic sensors. Adhesion at the interface of two materials can be controlled by various methods such as chemical
treatment on the surface of the materials, modification of the surface texture of materials, and change of the
mechanical properties of materials. The main idea of this study is to control the adhesion by changing the mechanical properties (modulus) of polydimethylsiloxane (PDMS) elastomer. We vary the modulus of PDMS elastomer by changing the silicone elastomer base's mixing ratio (w/w) and its curing agent (Sylgard$^\mathrm{TM}$ 184, Dow Corning). Our study also includes the effect of the thickness of the
PDMS elastomer sheet on its adhesion behavior. Adhesion measurements at the interface of the borosilicate glass slide and different PDMS elastomer specimens were performed using a wedge test. This method inserts a glass coverslip at the interface to create the wedge. We observe a significant decrease in the work of adhesion and an increase in equilibrium crack length with an increase in elastic-modulus and thickness of the PDMS elastomer samples. We present and discuss the effect of modulus and specimen-thickness on the adhesion behavior of PDMS elastomer against glass slide.
\end{abstract}

\maketitle

\section{Introduction}
Polymers are large molecules made of covalent-bonded repeating units known as monomers. They fall under the class of soft materials, with $k_\mathrm{BT}$ being the relevant energy scale. Mechanical and thermal properties of polymers, such as modulus, viscosity, melting temperature, etc., are strongly correlated with their molecular weight \cite{staudinger1920,r1}. For example, at low molecular weight, polymers behave like a Newtonian fluid, and with an increase in molecular weight, the behavior becomes non-Newtonian \cite{r3}. This is due to an increase in the number of entanglements in the system as molecular weight increases. The presence of entanglements also affects the glass transition temperature of polymers \cite{r4}. Due to the ease of processing, polymeric materials find applications in packaging, defense, electronic, {household} uses, etc. At times, it is important to tune the properties of polymers for specific applications. Crosslinking is one of the pathways to tune the properties of polymers for the desired applications \cite{singh2018combined, DMPRM21, Manoj_PRE, Manoj_macrolett, ain2024insights, Qurat_CI, Qurat_TI}. In the current work, we study the effect of crosslinking on the modulus of PDMS elastomer and, subsequently, the effect of modulus on the adhesion behavior of PDMS elastomer against glass slides. We also study the effect of the thickness of PDMS elastomer on their adhesion behavior. 

 It is important to highlight here that crosslinked polymer networks can be from weakly-crosslinked elastomers to highly-crosslinked epoxy. Therefore, by changing the ratio of cross-linkers to PDMS, the mechanical properties of the PDMS network can be tuned to make it significantly hard or soft. Following this idea, several studies have been conducted to enhance the mechanical properties of PDMS, such as modifying the elastic
modulus, hardness, stiffness, and adhesion at the interface \cite{carrillo2005nanoindentation,ALISAFAEI20131220}. Changes in crosslinking of silicone elastomer modulate the silicon networks and formulations that can produce soft, reproducible, and reliable elastomers \cite{stricher2015met,mazurek2019design}. For example, High-consistency rubber (HCR), thermoplastic silicone elastomers (TPE), and liquid silicone rubber (LSR) have different formulations. TPEs' mechanical behavior falls short of those of covalently bonded LSR and HCR elastomers. The presence of an uneven crosslink network in HCR shows viscous behavior, although LSR has a well-arranged network with high elasticity \cite{stricher2015met}. 

Adhesion is an attractive interaction between surfaces in contact. As a surface phenomenon, it especially becomes dominant at micro and nanoscales as the area-to-volume ratio increases. It is also a measure of the cause of failure in micro-electromechanical system (MEMS) and nano-electromechanical system (NEMS) devices \cite{bhushan2005adhesion,tambe2005micro}. 
As technological advancement leads us to miniaturization, it is important to control/tune the interfacial and mechanical properties of the material. Ghatak et al. study that when a thin elastic film is constrained and influenced by adhesion forces, it does not only homogeneously deform but rather undertake surface undulations \cite{ghatak2003adhesion}. These undulations create instability patterns in the system. They have studied the morphological characteristics of these instabilities and conducted contact mechanics tests comparable to the peel test in various geometries. Darby et al. study the three typical elastomer systems to produce a data set for interfacial and mechanical properties by changing the mixing ratios\cite{darby2022modulus}. These elastomers are Sylgard$^\mathrm{TM}$ 184, Solaris, and Ecoflex 00–30. Additionally, they have produced similarities in their adhesive features. They used parallel plate oscillatory shear rheology to quantify the shear storage moduli, spherical probe adhesion testing to evaluate the debonding work, and a goniometer to measure the contact angles. In general, softer samples are created by increasing the mixing ratios, and these samples are highly adhesive. Sylgard$^\mathrm{TM}$ 184 is a better adhesive than the Solaris, followed by Ecoflex 00–30. Nase et al. examine the debonding energy of the PDMS elastomer \cite{nase2013debonding}. In this study, they examine the effects of modulus, debonding velocity, and sample thickness on the process of debonding. They used the tack test geometry test to study the PDMS debonding process. This test has the benefit of having exact control over the debonding geometry, which makes it possible to directly examine the effects of thickness in restricted systems.
In numerous research, Sylgard$^\mathrm{TM}$ 184 PDMS elastomer was used as a model system for various mechanical and interfacial studies, e.g., adhesion, friction, wrinkling, etc.
Jagota et al. used Sylgard$^\mathrm{TM}$ 184 PDMS elastomer to study highly selective adhesion on wrinkle surfaces \cite{vajpayee2011adhesion}. Artz et al. also used Sylgard$^\mathrm{TM}$ 184 PDMS elastomer to study the adhesion behavior of PDMS to understand the gecko adhesion mechanism \cite{kroner2010adhesion,greiner2009experimental}. Wynne et al. used Peakforce Quantitative Nanomechanical Mapping (PF-QNM) using an atomic force microscopy (AFM) to study the AFM tip-PDMS surface interaction \cite{nair2019afm}. In their study, they also deduced the elastic modulus of Sylgard$^\mathrm{TM}$ 184 PDMS elastomer, which was in good agreement with independent bulk measurement. Stafford et al. developed a buckling-based metrology to quantify the elastic modulus of Sylgard$^\mathrm{TM}$ 184 PDMS elastomer \cite{stafford2004buckling}. Bonaccurso et al. studied static and dynamic wetting on soft and deformable Sylgard$^\mathrm{TM}$ 184 PDMS elastomer with a varying base to crosslinker ratio \cite{chen2011short,pericet2008effect}. Chakraborty et al. studied electrowetting on soft dielectric Sylgard$^\mathrm{TM}$ 184 PDMS elastomer films and found systematic variation in contact angle-voltage behavior based on base to crosslinker ratio of PDMS \cite{dey2017electrowetting}.
 PDMS is a material that is most commonly used to make dry adhesive elements such as gecko-inspired patterns and polymer brushes \cite{r27,r28,r29,r30}. The gecko is a creature with a highly developed hierarchical micro-structure that enables it to attain great adhesion on the majority of surfaces with varying degrees of roughness. Characteristics of this phenomenon have been examined in Ref \cite {r31}. using PDMS as the model substance since it can be structured using lithographic techniques and has exceptional mechanical properties \cite{r32}. PDMS is frequently used as the contacting layer in triboelectric nanogenerators for energy scavenging \cite{r33,r34}, where adhesion can play a crucial role in attaining a larger surface charge \cite{r35}. In the case of microfluidic devices, the adhesion of PDMS has great significance since it assures the proper contact between glass and PDMS to prevent leakage \cite{r36}.
Numerous works have been conducted to produce adhesive
polymeric surfaces based on chemical treatment of the surface \cite{r37,r38,r39}, by creating the heterogeneities on the surface or in bulk \cite{r41}. However, the control of the adhesion
by modifying the mechanical properties, such as the material's stiffness, has been much less studied. The main objective of this work is the adhesion
control at the interface of the borosilicate glass slide and the PDMS, which can be achieved by modifying the stiffness of the PDMS. The different stiffnesses of the PDMS can be achieved
by changing the mixing ratios of the PDMS base and the curing agent (w/w) and by varying the thickness of the PDMS sheet. We
conducted a wedge test to evaluate the adhesion at the interface of the PDMS against a borosilicate glass
slide. To create the wedge at the interface, we used a glass coverslip (thickness, $\delta \approx$ 0.15 mm).

\section{Material and methodology}
\subsection{Material}

PDMS (Polydimethylsiloxane) (make: Sylgard$^\mathrm{TM}$ 184, Dow Corning) is a silicon-based polymer. When PDMS is cross-linked, it transforms into a hydrophobic elastomer that can be cast to replicate different patterns with micro and nanoscale resolution \cite{r5,r6,r7}. PDMS is known for its extraordinary
properties and applications in different fields. The PDMS properties are optical
transparency, biocompatibility, exceptional mechanical properties, chemical stability, and excellent
rheological properties \cite{r8,r9,r10,r11,r12}. Due to these properties, PDMS has broad applications in the engineering
and biomedical fields, such as microvalves \cite{r13}, micropumps \cite{r14}, microfluidics and photonics \cite{r15,r16,r17,r18},
dressings and bandages \cite{r19}, the study of diseases \cite{r20,r21}, and soft-lithography \cite{r16,r17,r22}. Besides these applications, PDMS also comprises contact lenses \cite{r23}, water-repellent coatings \cite{r24}, cosmetics \cite{r25}, lubricants \cite{r26}, and many more.

Conventionally, PDMS polymer refers to silicone oil, whereas PDMS elastomer refers to crosslinked PDMS. Sylgard$^\mathrm{TM}$ 184 is a common commercially available silicone elastomer from Dow, USA. For convenience, we referred to Sylgard$^\mathrm{TM}$ 184 silicone elastomer as PDMS in the current manuscript.

\subsection{Methodology}
\subsubsection{Sample preparation}

The following processes were used to make the two different sets of samples. At first, we made the PDMS sheets having different mixing ratios base and curing agents. The PDMS base and curing agent (Sylgard$^\mathrm{TM}$ 184, Dow Corning) were mixed in w/w ratio of 5:1, 10:1, 15:1, and 20:1. For better mixing, the mixtures were stirred in a beaker with a glass rod. During the mixing process, air bubbles were formed. To eliminate the air bubbles, each mixture was kept in the desiccator at a pressure below ambient for 20 min. Then, the mixture was poured onto the glass plate, and another glass plate was used to cover it. To fix the thickness of the PDMS sheet, we used the 1 mm thick flat washer of stainless steel between the glass plates. Then, the mixture was cured at room temperature for an hour. Subsequently, the mixture was cured at the temperature of 120 $^oC$ for $2$ hr in the furnace, following which it was cooled to room temperature. Then, the PDMS elastomers formed were peeled off from the glass plate and cut into the sheets of dimensions 4 cm$\times$1 cm having a thickness $\approx$ 1 mm. Subsequently, the PDMS elastomers were used for adhesion tests. For the second set, we made the PDMS elastomers of different thicknesses by fixing the mixing ratio to 10:1. To control the thickness of the PDMS sheet. We used the flat washers of the stainless steel of various thicknesses ($\approx$ 0.45 mm, 0.75 mm, 1 mm, 1.35 mm). All other processes were the same as the ones mentioned above for the first set of samples with varying mixing ratios.

\subsubsection{Quasi-static nanoindentation test}

We expected a change in the modulus of elasticity with a change in the mixing ratio of the PDMS base and curing agent. The change in elastic modulus of the PDMS elastomer was examined by the quasi-static nanoindentation (Hysitron TI 750) test using the Berkovich tip \cite{r42}. The samples of the different mixing ratios were tested at different loads but at the same depth of indentation. The PDMS sheet thickness $\approx$ 3 mm was cut into the square pads of the area 1 cm$^2$. It is important to note that \textcolor{blue}{separate} set of 3 mm thickness samples were prepared for indentation experiments. We used glue (Fevikwik, Pidilite Industries Ltd.) at the interface of the sample pad and the nanoindenter stage to attach them and pressed gently with tweezers to confirm complete contact. 
For the nanoindentation using the Berkovich tip, the reduced elastic modulus of the sample can be calculated as follows:
\begin{equation} \label{eq_(1)}
E_r = \frac{S\sqrt{\pi}}{2\sqrt{A}}
\end{equation}
where $A$ is the contact area of the Berkovich tip. $S$ is the slope of the top part of the unloading curve. And for the Berkovich tip, $A$ can be given as follows:

\begin{equation} \label{eq_(2)}
A = 24.56 h^2
\end{equation}
where $h$ is the depth of the indentation. The relation between the elastic modulus of PDMS and the reduced modulus is given as follows:

\begin{equation} \label{eq_(3)}
\frac{1}{E_r} = \frac{(1 - \nu_{\mathrm{PDMS}}^2)}{E_{\mathrm{PDMS}}} + \frac{(1 - \nu_{\mathrm{tip}}^2)}{E_{\mathrm{tip}}}
\end{equation}

where $\nu_\mathrm{PDMS}$ is the Poisson's ratio of the PDMS. ($\nu_\mathrm{tip} = 0.07$) and ($E_{\mathrm{tip}} = 1140 \mathrm{GPa}$) are the Poisson's ratio and the elastic modulus of the diamond indenter, respectively \cite{r42}.

Since the elastic modulus of the diamond indenter is much higher than the elastic modulus of the PDMS, the second term of the equation \eqref{eq_(3)} does not make a significant contribution. So, the final relation between the elastic modulus of the PDMS and the reduced modulus is given as follows:

\begin{equation} \label{eq_(4)}
E_{\mathrm{PDMS}} = (1 - \nu_{\mathrm{PDMS}}^2) E_r
\end{equation}

\subsubsection{Wedge test}
An inverted microscope (IX 73, Olympus, Japan) was used to examine the crack propagation at the interface of PDMS elastomer and glass slide. We first fixed a glass slide on the inverted microscope stage and clung a PDMS sheet over it. A thin glass coverslip of the thickness ($\approx$ 0.15 mm) was inserted at the interface of the PDMS elastomer specimen and glass slide to create the wedge. The PDMS elastomer sheet was pressed at the interface with a smooth glass rod of diameter ($\approx 5 $ mm). It is necessary to apply adequate pressure at the interface to attain good contact. After the release of pressure, the crack started propagating due to a spontaneous opening at the edge of a glass coverslip. The crack propagation was recorded by commercial software, view7. The recorded videos were analyzed by the open-source software Tracker \cite{c22_brown2009innovative,c23_moraru2021distance,c24_rodrigues2013teaching,c25_wee2012using}. The growth in crack length of samples having different moduli and thicknesses was examined as a function of time. The equilibrium crack length (where the crack propagation stops) \cite{r41} was noted for each PDMS elastomer specimen.
\begin{figure}[h]
\centering
  \includegraphics[height=5.5cm]{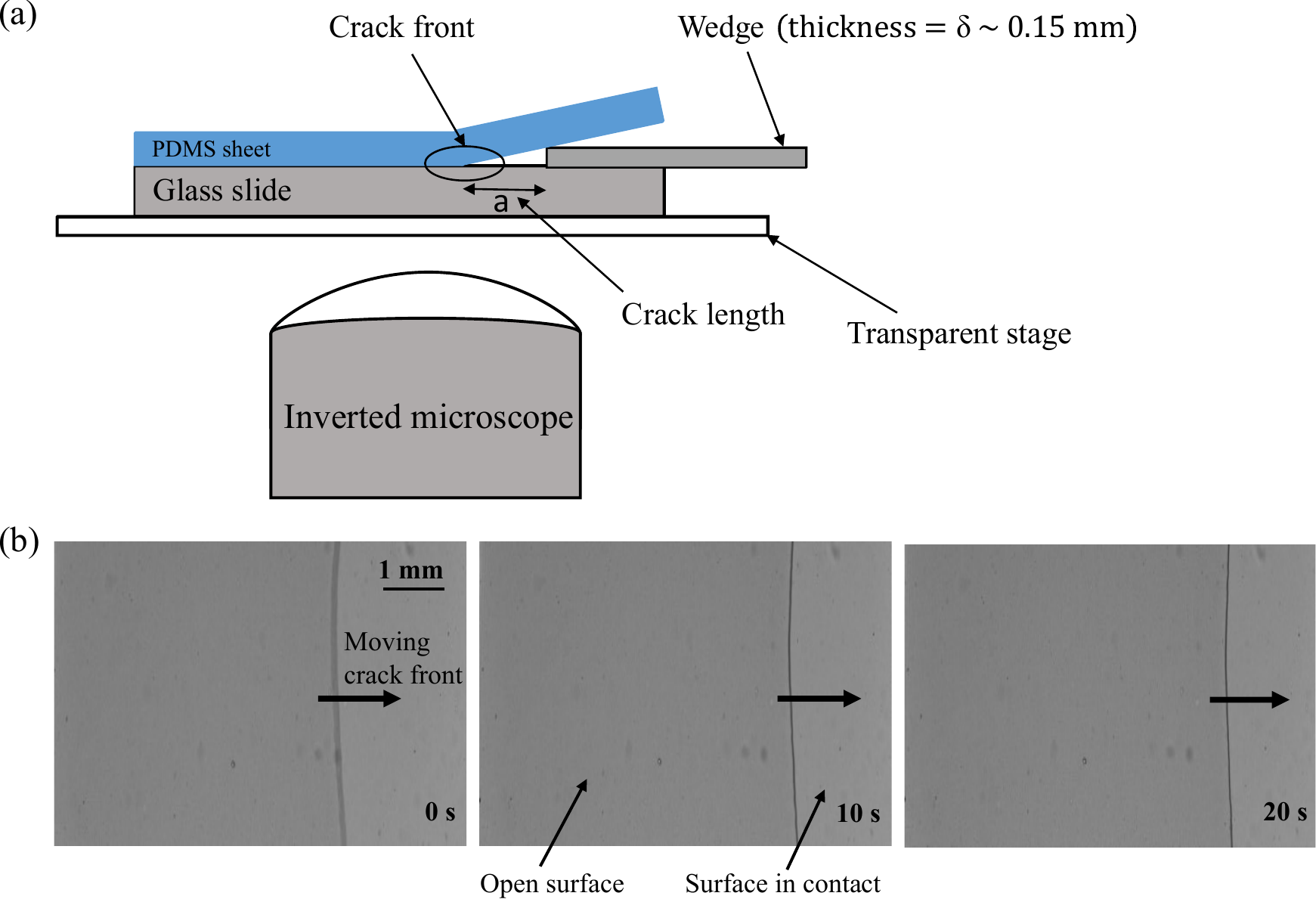}
  \caption{(a) Schematic of the experimental setup built in the inverted optical microscope (IX 73, Olympus, Japan) (b) Optical image of the crack propagation captured at different periods. Where the head of the arrow points to the surface in contact, and the tail of the arrow points to the open surface.}
  \label{fgr:exp_setp}
\end{figure}

Figure \ref{fgr:exp_setp} (a) shows a schematic of the experimental setup. As shown in the figure, a high-speed camera (up to 100 fps) was used with the inverted microscope. The figure also shows the glass slide, PDMS-elastomer sample, and the wedge inserted at their interface. The crack propagation at the interface was recorded using the high-speed camera as a function of time. Fig.\ref{fgr:exp_setp} (b) shows a sample of optical images of the crack propagation captured using the high-speed camera at different periods. PDMS sheets of different elastic modulus (obtained by varying the mixing ratio of Sylgard base and crosslinker) and different thicknesses were used to study their effect on the adhesion between PDMS elastomer and glass slides.

\section{Results and discusion}
\begin{figure}[h]
    \centering
    \includegraphics[width=0.49\textwidth,angle=0]{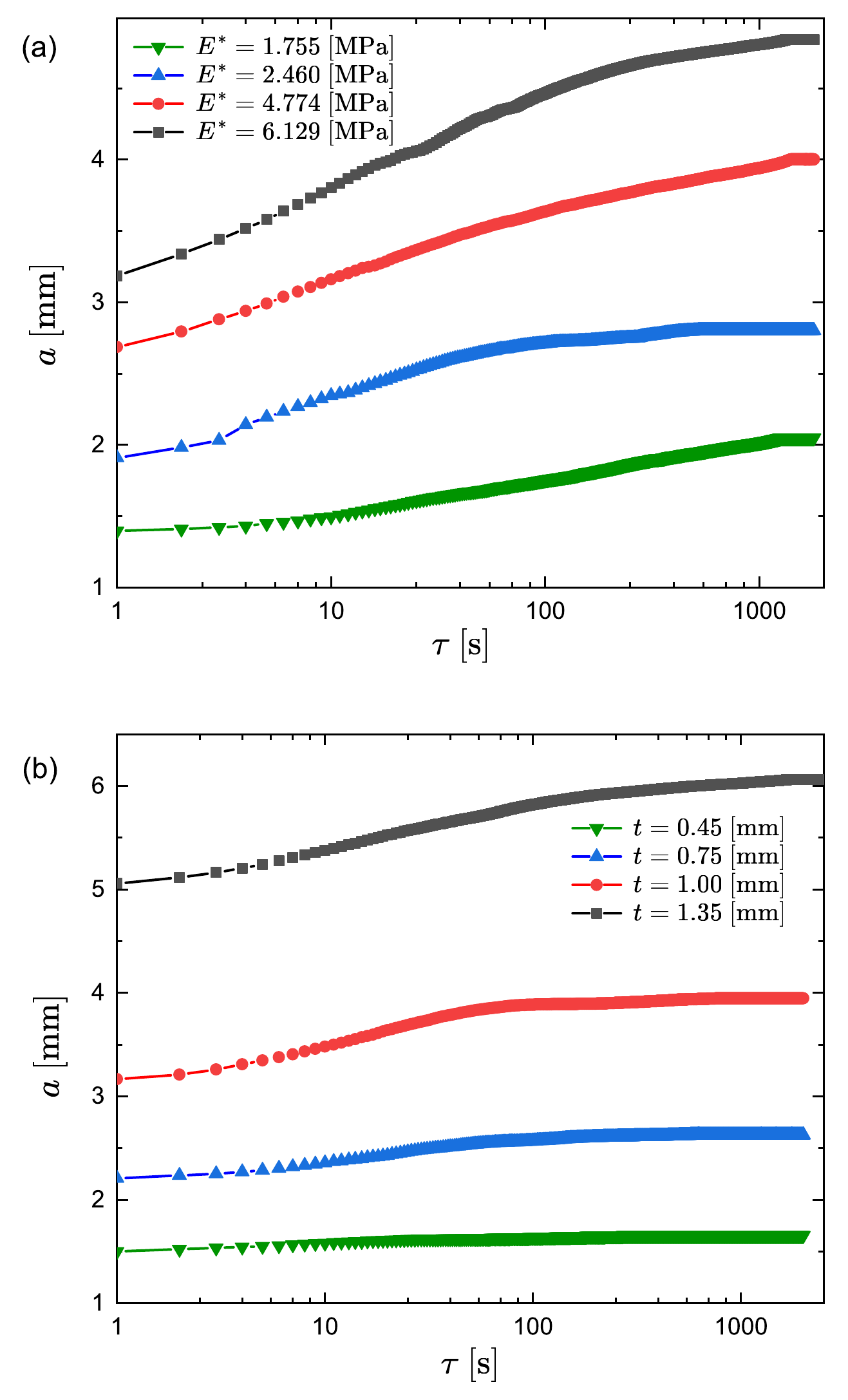}
    \caption{Crack length $(a)$ versus time $(\tau)$ plot. (a) Each curve is related to the different values of the $E^*$ and shows a sufficient increase in the crack length with an increase in the $E^*$. (b) Each curve is related to the different sets of data on the thickness $(t)$ of the PDMS elastomer and shows an adequate increase in the crack length with an increase in the $t$ of the PDMS elastomer.}
    \label{fig:ink_at}
\end{figure}

 Mode I crack propagation theory has been used in the current study to understand the effect of the modulus and thickness of PDMS sheets on their adhesion behavior. Adhesion was estimated through crack propagation at the interface of the PDMS elastomer samples and glass slide \cite{r50}. We have taken the smooth surface assumption for the samples, as the roughness is negligible in comparison to other dimensions, such as sample size and crack length. Hence, we neglect the effect of roughness at the interface and focus on a semi-infinite crack tip subjected to remote mode I loading. In the case of elastic material, when a normal tensile load is applied, the crack propagates perpendicularly to the direction of loading as the energy release rate is maximum in that direction. For a flat interface without any waviness, this is the same direction for crack propagation. Even so, the direction of crack propagation may deviate for an interface with poor wavy contact. So, across the crack path, there will be some point where the energy release rate in a flat interface without waviness will be more than the wavy interface. It can be illustrated by calculating the maximum energy release rate when it makes an angle $\theta$ from the direction of crack propagation. If the open crack face behind the crack tip is taken as flat, then the stress intensity factor can be given by \cite{r50}
\begin{equation} \label{eq_(5)}
K^{'}_I = K_I f^I_{\mathrm{\theta\theta}}
\end{equation}
\begin{equation} \label{eq_(6)}
K^{'}_{\mathrm{II}} = K_I f^I_{\mathrm{r\theta}}
\end{equation}
where $K_I$ is the mode I stress intensity factor and $K^{'}_{\mathrm{II}}$ is the mode II stress intensity factor. $f^I_{\mathrm{\theta\theta}}$ and $f^I_{\mathrm{r\theta}}$ can be given by:
\begin{equation} \label{eq_(7)}
f^I_{\mathrm{\theta\theta}} = \cos^3{({\theta}/2)}
\end{equation}
\begin{equation} \label{eq_(8)}
f^I_{\mathrm{r\theta}} = \sin({\theta}/2)\cos^2({\theta}/2)
\end{equation}
After combining both equations, the energy release rate $G_r$ can be given by:
\begin{equation} \label{eq_(9)}
G_r = \frac{K^{'2}_I + K^{'2}_{\mathrm{II}}}{E^*}
\end{equation}
\begin{equation} \label{eq_(10)}
G_r = \frac{K^2_I}{E^*} \cos^4({\theta}/2)
\end{equation}

where $E^* = \frac{E}{(1-\nu{^2})}$ is the plane strain modulus, $E$ is the young modulus of elasticity and $\nu$ is the poison’s ratio of the PDMS elastomer. From the equation \eqref{eq_(10)}, it is clear that the energy release rate is decreased by a factor $\cos^4(\frac{\theta}{2})$. For the maximum value of $\theta$, the energy release rate will be minimum along the wavy surface. And can be given by:

\begin{equation} \label{eq_(11)}
G_r = W_{\mathrm{ad}} \cos^4({\theta_{\mathrm{max}}}/2)
\end{equation}
where ${W_{\mathrm{ad}}} =  \frac{K^2_I}{E^*}$ is the work of adhesion. According to Waters and Guduru, under mixed loading conditions, the effective work of adhesion increases at the interface \cite{r44}. So energy release rate can be given as follows:    
\begin{equation} \label{eq_(12)}
G_r = W^r_{\mathrm{ad}}({\Psi},\upsilon)
\end{equation}
where $W^r_{\mathrm{ad}}$ is the work of adhesion at the interface, $\Psi$ is the phase angle and $\upsilon$ is the crack velocity. $\tan{\Psi} = \frac{K'_{\mathrm{II}}}{K'_I} = \tan({\theta}/2) \implies \Psi = (\frac{\theta}{2})$, Putting this value of the phase angle in the above equation, we get
\begin{equation} \label{eq_(13)}
G_r = W^r_{\mathrm{ad}}({\theta}/2,\upsilon)
\end{equation}
From the equation \eqref{eq_(13)}, it seems that in a quasi-static crack growth, the energy release rate for a flat contact is related to the work of adhesion by:
\begin{equation} \label{eq_(14)}
G_r = W^r_{\mathrm{ad}}({\theta} = 0,\upsilon) = W_{\mathrm{ad}}(\upsilon)
\end{equation}
$W^0_{\mathrm{ad}} = W_{\mathrm{ad}} (\upsilon=0)$ is the threshold value of work of adhesion for a flat interface. It is found when the crack velocity approaches zero. But for a definite value of velocity, $W_{\mathrm{ad}} (\upsilon) > W^0_{\mathrm{ad}}$. when the crack length increases, the crack velocity goes down, and as a result, the energy release rate $G_r$ reduces. So, the point of time when crack growth stops is the equilibrium point where the equilibrium crack length will be $a= a_e$. And this is the point when the $W_{\mathrm{ad}} (\upsilon)$ approaches a constant value of $W^0_{\mathrm{ad}}$. From equations \eqref{eq_(10)} and \eqref{eq_(14)}, we can write
\begin{equation} \label{eq_(15)}
W_{\mathrm{ad}} = \frac{K_I^2}{E/(1-\nu^2)}
\end{equation}
For a fixed end or constant displacement $(u)$, the energy release rate $G_r$ for mode I can be given by:
\begin{equation} \label{eq_(16)}
G_r = \frac{3E^*\delta^2t^3}{4a^4}
\end{equation}
where $u=2\delta$ ($\delta$ = thickness of glass coverslip), $t$ is the thickness of the PDMS elastomer sample and $a$ is the crack length. From Fig.\ref{fig:ink_at}, it is clear that initially, the crack length proliferates since the energy release rate is more significant for lower crack lengths. Then, it slows down since the energy release rate is lower for higher crack lengths (later discussed in detail). So from equations \eqref{eq_(14)} and \eqref{eq_(16)}, we can show the relation between crack length and work of adhesion as follows:
\begin{equation} \label{eq_(17)}
W_{\mathrm{ad}} = \frac{3E^*\delta^2t^3}{4a^4} = \frac{K_I^2}{E^*}
\end{equation}
And at the equilibrium point, where the crack length is $a=a_e$. 
\begin{equation} \label{eq_(18)}
W^0_{\mathrm{ad}} = \frac{3E^*\delta^2t^3}{4a_e^4}
\end{equation}
Thus, the equation \eqref{eq_(18)} shows the relation between the work of adhesion and the equilibrium crack length at the interface of the PDMS elastomer and glass slide.

It should be made clear that $W^0_\mathrm{ad}$ is better understood as fracture energy, which comes from irreversible and reversible processes and possesses the same units as surface energy. However, $W^0_\mathrm{ad}$ should not be treated as surface energy or thermodynamic work of adhesion since these are likely to stay the same with crack speed/crack length and have less possibility of being as large as $W^0_\mathrm{ad}$ \cite{kendall1975thin}.  

\begin{figure}[h]
    \centering
    \includegraphics[width=0.49\textwidth,angle=0]{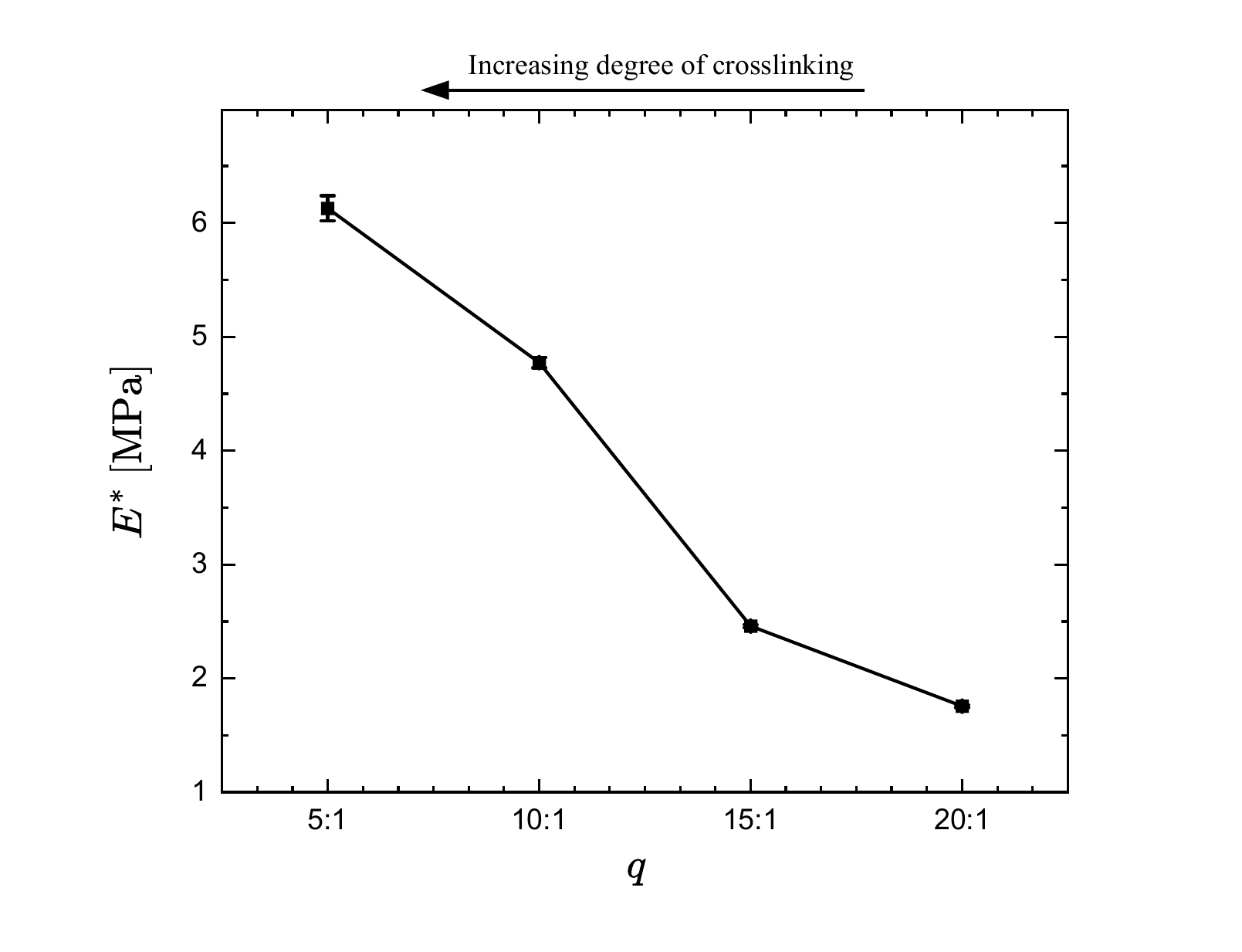}
    \caption{Shows the plot of \textcolor{blue}{plane} strain modulus $(E^*)$ versus mixing ratio \textcolor{blue}{$(q)$} in w/w ratio of the PDMS base and the curing agent. The error bar is included in the plot, which was found to be $(<2\%)$.}
    \label{fig:E_mr_nano indentation}
\end{figure}

Figure \ref{fig:E_mr_nano indentation} shows the plot of \textcolor{blue}{plane} strain modulus ($E^*$) versus the mixing ratio of the PDMS base and the curing agent. The symbol $q$ shows the mixing ratio (w/w). The reduced elastic modulus of the PDMS elastomer of the different mixing ratios was found by the quasi-static nanoindentation test (based on Oliver and Pharr analysis) \cite{oliver2004measurement} using the Berkovich tip \cite{r42}. For the calculation of the \textcolor{blue}{plane} strain modulus of the PDMS elastomer, the poison ratio of the PDMS elastomer was taken as $(\nu = 0.5)$ \cite{pr_1} as at room temperature PDMS elastomer can be assumed linearly elastic and incompressible \cite{pr_2,pr_21}. In the test, it is found that, as the concentration of the curing agent diminished in the mixture of the PDMS base and the curing agent, the elastic modulus of the PDMS elastomer reduces, and consequently, the \textcolor{blue}{plane} strain modulus of the PDMS elastomer decreases. Our findings are \textcolor{blue}{consistent} with recently published work \cite{Manoj_PRE}.

Figure \ref{fig:ink_at} highlights the plots of the crack length versus time for samples of different \textcolor{blue}{plane} strain modulus and thicknesses of PDMS elastomer. Data of the crack length was extracted from Fig. \ref{fgr:exp_setp} (b) corresponding to each sample. It can be seen that in all the samples at first, the crack length increases rapidly with an increase in time. As discussed earlier, the initial displacement ($u=2\delta$) in all the samples is the same as the thickness of the wedge created by the coverslip glass. At the time when crack propagation stops, we get the threshold value of crack length (equilibrium crack length, $a_e$). Once the equilibrium crack length is reached, crack propagation velocity becomes zero, and the corresponding value of the energy release rate is the threshold value, i.e., work of adhesion. The energy release rate is responsible for crack propagation, decreasing as crack length increases. Also, the crack propagation velocity decreases with an increase in crack length. The increment in the crack length is nonuniform since the crack propagation velocity is changing non-uniformly. We also observed that with a decrease in modulus of elasticity, the equilibrium crack length reduces, and thus, the threshold value of the energy release rate increases. Similarly, when the thickness of the PDMS sheet decreases, the equilibrium crack length decreases; thus, the threshold value of the energy release rate increases. These observations are discussed below in detail.
\begin{figure}[h]
    \centering
    \includegraphics[height = 13cm]{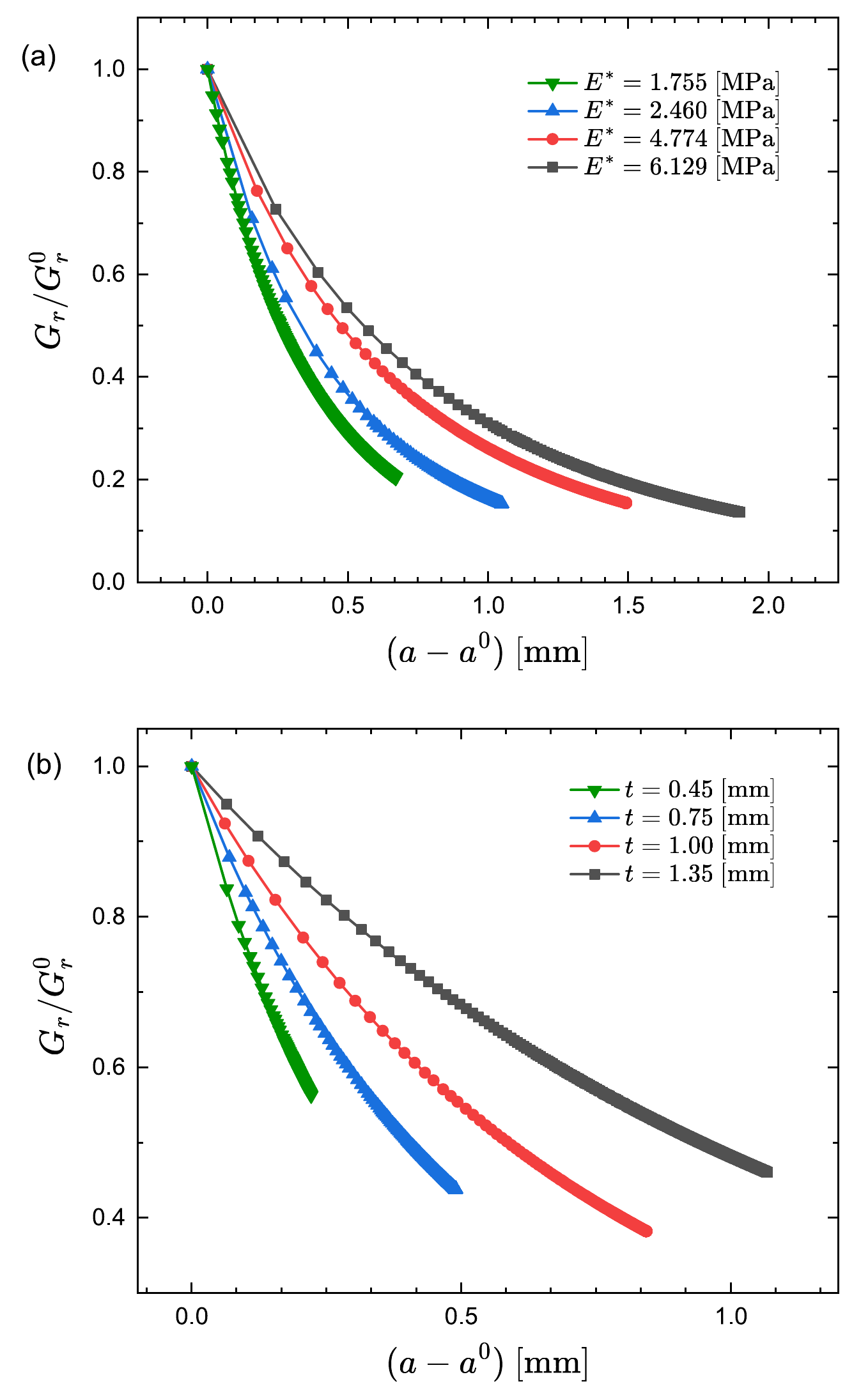}
    \caption{(a) Shows the plot of normalized energy release rate $({G_r}/{G_r^0})$ versus crack length $(a-a^0)$ for the different values of $E^*$. The normalized energy release rate is the ratio of the energy release rate to the initial value of the energy release rate at $(\tau=0\space\mathrm{s})$. $a^0$ is the initial crack length at $(\tau=0\space\mathrm{s})$. (b) Shows the plot of normalized energy release rate versus crack length for the different values of $t$ of the PDMS elastomer.}
    \label{fig:ink_Ga}
\end{figure}
Fig. \ref{fig:ink_Ga} illustrates the plot of the normalized energy released rate versus crack length for different values of the \textcolor{blue}{plane} strain modulus and the thickness of the PDMS elastomer. The normalized energy release rate is the ratio of the energy release rate ($G_r$) to the initial value of the energy release rate ($G_r^0$) at time $\tau = 0 \space\mathrm{s}$. $a^0$ is the initial crack length at $\tau=0 \space\mathrm{s}$. The normalized energy release rate reduces with an increase in the crack length in both the cases of \textcolor{blue}{plane} strain modulus and thickness of the PDMS elastomer. The threshold value of the crack length in both cases sufficiently increases with an increase in \textcolor{blue}{plane} strain modulus and thickness of the PDMS elastomer, respectively. The rate of reduction of the normalized energy release rate with a change in crack length 
$(a-a^0)$ adequately increases with a decrease in the \textcolor{blue}{plane} strain modulus and also a decrease in the thickness of the PDMS elastomer. Due to that reason, the energy release rate reduces sharply with an increase in the crack length for a smaller value of the \textcolor{blue}{plane} strain modulus and the thickness of the PDMS elastomer.  

\begin{figure}[h]
    \centering
    \includegraphics[width=0.49\textwidth,angle=0]{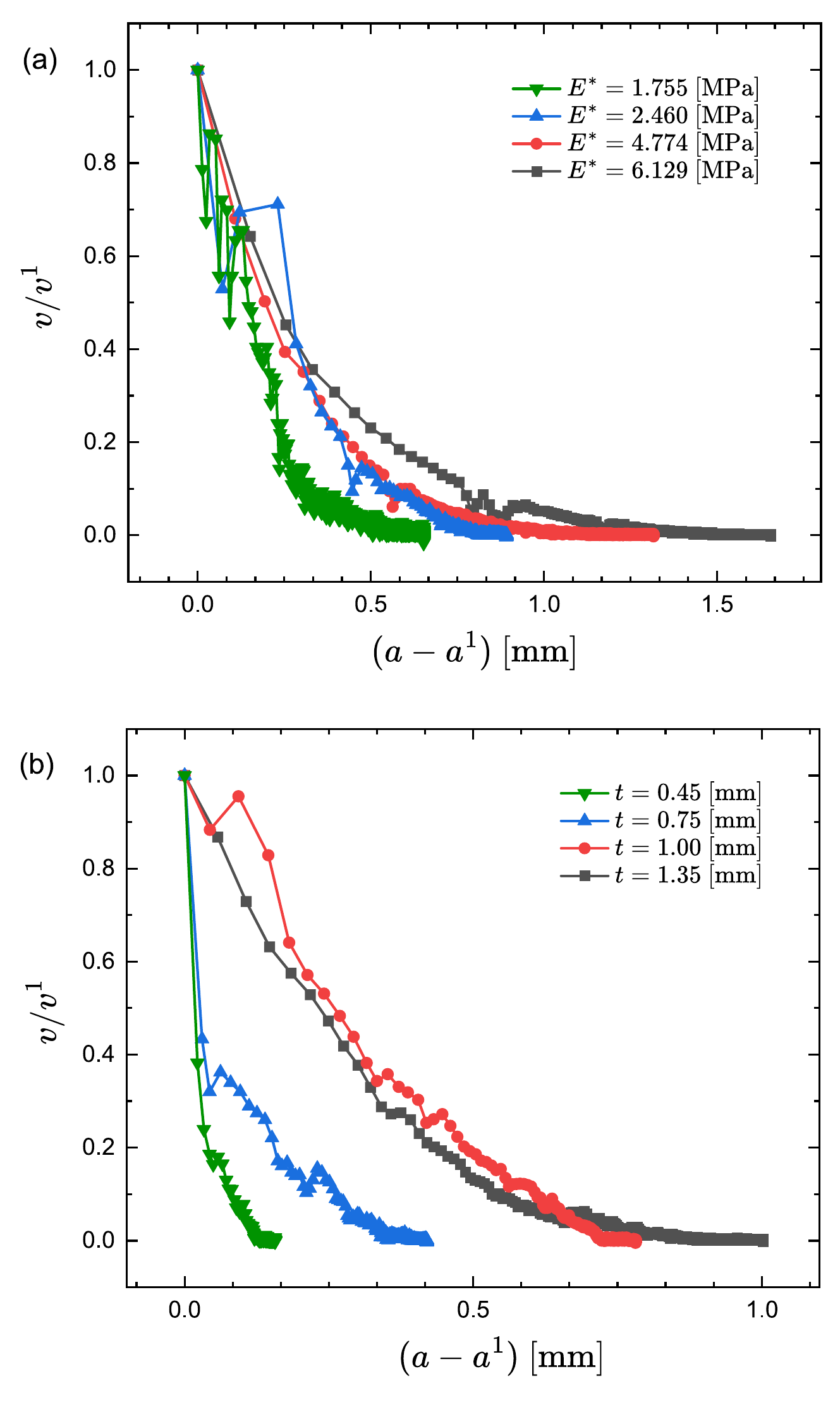}
    \caption{(a) Shows the plot of normalized crack velocity \textcolor{blue}{$(v/v^1)$} versus crack length \textcolor{blue}{$(a-a^1)$} for the different values of $E^*$. The normalized crack velocity is the ratio of the crack velocity to the crack velocity at time blue{$\tau=1\space\mathrm{s}$}. (b) Shows the plot of normalized crack velocity versus crack length for the different values of $t$ of the PDMS elastomer.}
    \label{fig:ink_va}
\end{figure}

Figure \ref{fig:ink_va} shows the plot of normalized crack velocity versus the crack length for the different values of the plane strain modulus and the thickness of the PDMS elastomer. The normalized crack velocity is the ratio of the crack velocity ($v$) to the crack velocity ($v^1$) at time $\tau=1\space\mathrm{s}$ (we have chosen $\tau=0\space\mathrm{s}$ as the initial point for observing crack propagation. At this point, the crack velocity is zero, although the crack length acquires a finite value. At $\tau=1\space\mathrm{s}$, both crack velocity and crack length reach a finite value. So, we have normalized our crack velocity from $\tau=1\space\mathrm{s}$). The normalized crack velocity was decreased with an increase in the crack length in both the cases of plane strain modulus and thickness of PDMS elastomer. The reduction in normalized crack velocity with a change in crack length ($a-a^1$) is nonuniform. This could be due to dust particles at the interface of the PDMS elastomer and the glass slide. As discussed earlier, the threshold value of the crack length is found when the crack velocity approaches zero.
The rate of reduction of the normalized crack velocity with the crack length sufficiently increases with a decrease in the plane strain modulus and thickness of the PDMS elastomer. This can be confirmed by the fact that normalized crack velocity reduces sharply with an increase in the crack length for the smaller values of the plane strain modulus and the thickness of the PDMS elastomer. 

\begin{figure}[h]
    \centering
    \includegraphics[width=0.49\textwidth,angle=0]{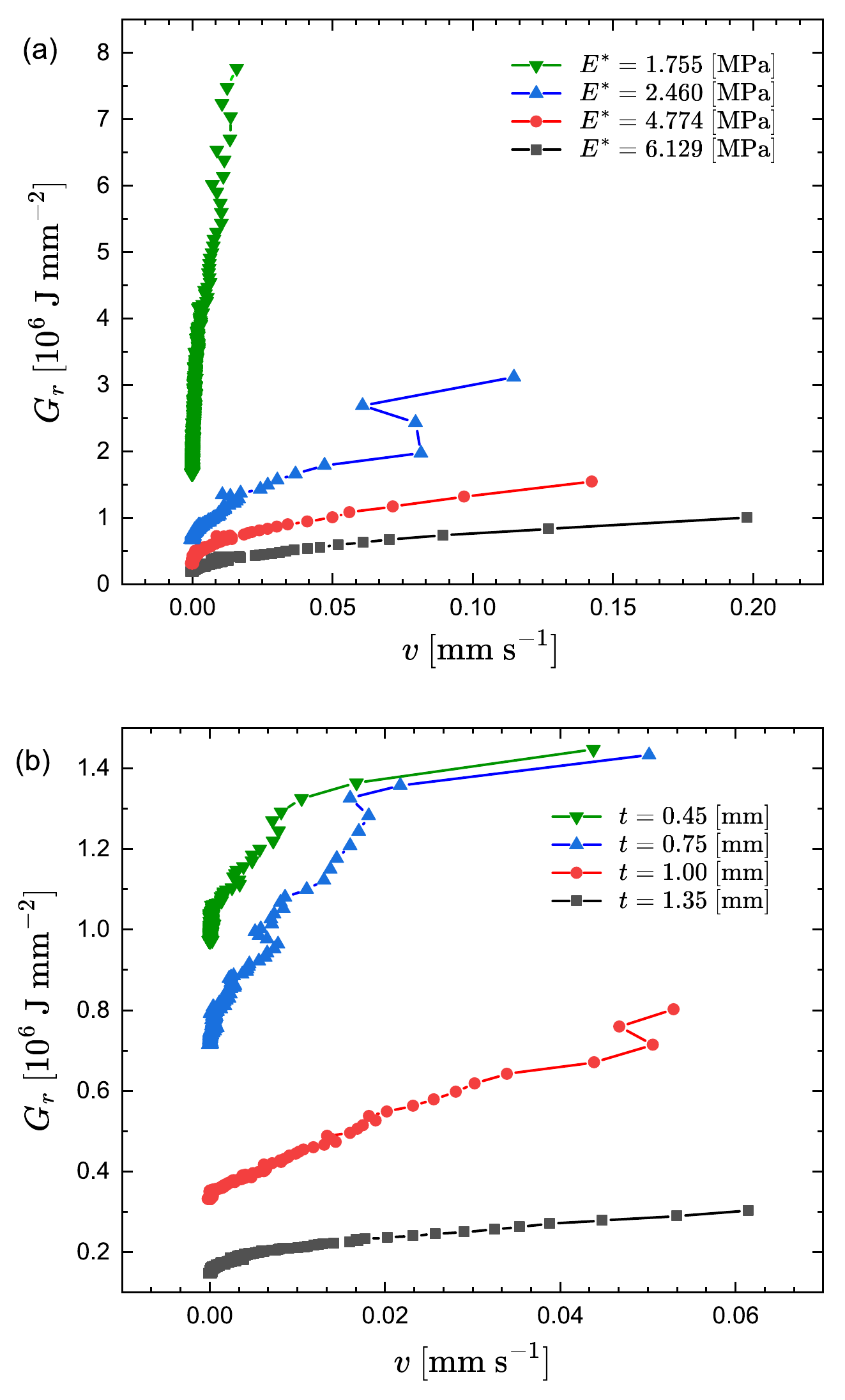}
    \caption{Plot of energy release rate $(G_r)$ versus crack velocity $(v)$ (a) Each curve is related to the different values of $E^*$ and shows an adequate increase in the threshold value of the energy release rate with a decrease in the $E^*$.(b) Each curve is related to a different $t$ of the PDMS elastomer and shows an adequate increase in the threshold value of the energy release rate with a decrease in the $t$ of the PDMS elastomer.}
    \label{fig:ink_Gv}
\end{figure}

Figure \ref{fig:ink_Gv} illustrates the plots of energy release rate ($G_r$) versus crack velocity ($v$) for the sample of different plane strain modulus and thicknesses of the PDMS elastomer. As already discussed in the raw data of the crack length versus time plot Fig.\ref{fig:ink_at}, advancement of crack growth stops when it reaches a specific crack length, i.e., equilibrium crack length ($a_e$). At $a_e$, the threshold value of the energy release rate (work of adhesion, $W_\mathrm{ad}^0$) is found, and the crack velocity ($v$) becomes zero. Furthermore, we observed that a decrease in the modulus and thickness of the PDMS elastomer leads to an increase in the $W_\mathrm{ad}^0$ (threshold value of the energy release rate) at the interface. 

Adhesion depends on the crack propagation velocity for all the interfaces. The cause of velocity dependency of adhesion at the interface of flat surfaces has been thoroughly investigated and comprehended \cite{r45,r46,r47,r48,r49,r50}.
 The velocity dependency of adhesion can be caused by the chemical reaction at the interface \cite{r47,r51,r52} or by viscoelastic energy loss in the bulk \cite{r45,r46,r48,r49,r50}. We have used PDMS elastomer in our study, which is an elastic material. Thus, the majority of the rate effects are most probably caused by kinetic interfacial processes rather than bulk viscoelasticity \cite{r46,r47,r48,r49,r50,r51,r52,r53}.

 \begin{figure}[ht]
    \centering
    \includegraphics[width=0.49\textwidth,angle=0]{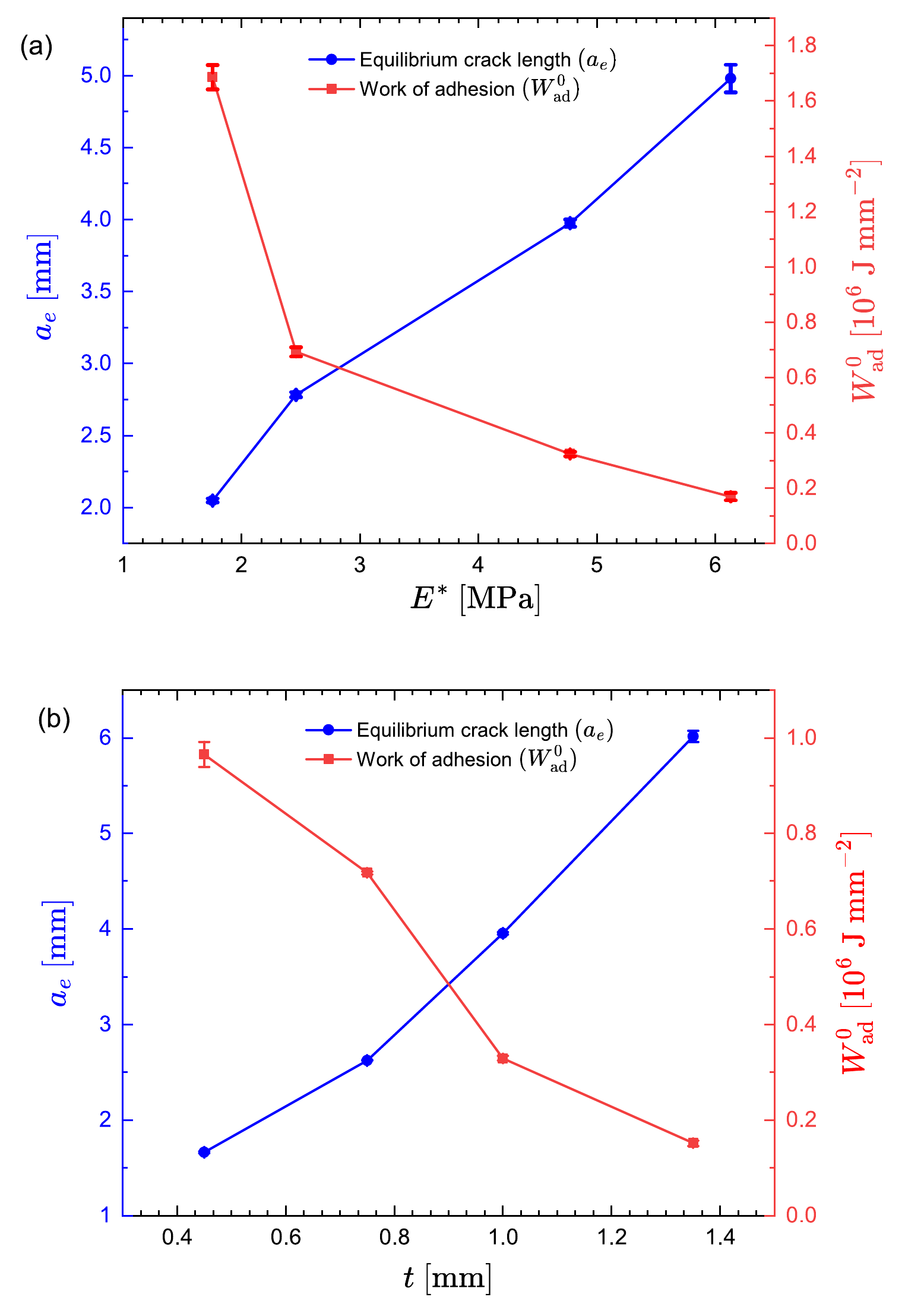}
    \caption{Illustration of the equilibrium crack length $(a_e)$ and the work of adhesion $(W^0_\mathrm{ad})$ corresponding to $E^*$ and $t$ of the PDMS elastomer. (a) Each curve signifies the data of equilibrium crack length and work of adhesion corresponding to the $E^*$ and shows an increase in the equilibrium crack length and a decrease in the work of adhesion with an increase in $E^*$. (b) Each curve is related to the equilibrium crack length and work of adhesion corresponding to the $t$ of the PDMS elastomer and shows an increase in the equilibrium crack length and a decrease in the work of adhesion with an increase in the $t$ of the PDMS elastomer.}
    \label{fig:ink_a0_Wad}
\end{figure}

Figure \ref{fig:ink_a0_Wad} presents the concluding results of the current study that includes plots of equilibrium crack length ($a_e$) and work of adhesion ($W_\mathrm{ad}^0$) versus the plane strain modulus ($E^*$) and thicknesses ($t$) of PDMS elastomer. As discussed earlier, the crack length corresponding to the zero value of the crack velocity is the threshold value of the crack length (equilibrium crack length, $a_e$), and the correspondent value of the energy release is the interfacial threshold value of the energy release rate (work of adhesion, $W_\mathrm{ad}^0$). The value of $a_e$ reduces with a decrease in both the modulus ($E^*$) and thickness ($t$) of the PDMS elastomer. The interfacial $W_\mathrm{ad}^0$ was calculated by substituting $a_e$ in the equation \eqref{eq_(18)}. A lower value of $a_e$ at the interface gives a higher value of $W_\mathrm{ad}^0$. Thus $W_\mathrm{ad}^0$ increases with a decrease in the concentration of the curing agent in the mixture of the PDMS base and the curing agent. $W_\mathrm{ad}^0$ also increases with a decrease in the thickness of the PDMS elastomer sheet. Hence, this study presented that the interfacial adhesion behavior of PDMS elastomer strongly correlates with modulus ($E^*$) as well as thickness ($t$) of the elastomer. This can be understood as follows: as the modulus of PDMS elastomer decreases, it makes more conformal contact with the glass slide. It leads to an increase in the real area of contact. Though the atomic-scale interaction between the materials remains unchanged, the increase in the real contact area is manifested in terms of an increase in adhesion between the interfaces. 

\section{Conclusions}
The current study shows that the adhesion at the interface of the PDMS elastomer and glass slide can be controlled by varying the concentration of the curing agent and thickness of the PDMS elastomer.  The energy release rate strongly correlates with the elastic modulus and thickness of the elastomer.

To study the effect of modulus, we varied the amount of curing agent in the mixture of the PDMS base and the curing agent from 5:1 to 20:1 systematically. The modulus of elasticity was measured using nanoindentation, and we found that the elastic modulus of PDMS elastomer reduces on decreasing content of curing agent, i.e., PDMS elastomer became softer. From our wedge test, we observed that equilibrium crack length ($a_e$) increases and work of adhesion ($W_\mathrm{ad}^0$) decreases with increase in modulus ($E^*$). It can be explained as follows: when a softer PDMS sheet is pressed against the glass slide, the PDMS elastomer makes more conformal contact with the glass slide than the PDMS elastomer with a higher modulus. The conformal contact leads to an increase in the real area of contact, which is reflected in the form of an increase in adhesion at the interface and a decrease in the equilibrium crack length. 

We also varied the thickness ($t$) of the elastomer specimen while keeping the mixing ratio of PDMS base and curing agent constant at 10:1. Here again, an increase in equilibrium crack length and a decrease in work of adhesion was observed with an increase in thickness of PDMS specimens. This can be understood in terms of the bending of beams: as the $t$ increases, for identical deflection ($\delta \approx 0.15$ mm), beam length, i.e., equilibrium crack-length ($a_e$), has to be higher.  We also observed in both cases (varying $t$ and $E^*$) that the energy release rate ($G_r$) decreases with the increase in the crack length. Initially, it reduces fast, but after some time, when the crack length approaches the equilibrium value, and the velocity of crack propagation approaches zero, then $G_r$ reaches its threshold or the constant value. The velocity of crack propagation also decreases with the increase in the crack length. Initially, it decreases rapidly since the energy release rate is high. It approaches zero after some time since the energy release rate reaches the threshold value. The energy release rate decreases with a decrease in crack propagation velocity. When velocity approaches zero, the energy release rate reaches the threshold value. 

This work will be useful in designing elastomer-based micro-channels and other miniature components with desired adhesive properties. The approach employed in the current study can be taken forward to incorporate the effect of surface-roughness and thermal-ageing on the adhesion behavior of elastomers.

\section*{Conflicts of interest}

The authors declare no conflict of interest.

\section*{Data Availability Statement}

The data presented in this work will be shared on reseaonable request.

\section*{Acknowledgements}
M.K.S. thanks the Science and Engineering Research Board (SERB), India, for the financial support provided under the Start-up Research Grant (SRG) scheme (Grant No. SRG/2020/000938). K.K. acknowledges the funding support from SERB, New Delhi (Project no. CRG/2019/000915).

\bibliographystyle{ieeetr}
\bibliography{reference}
\end{document}